\documentstyle[epsfig]{mn}
\DeclareGraphicsExtensions{ps}
\def\be{\begin{equation}}
\def\ee{\end{equation}}
\def\bea{\begin{eqnarray}}
\def\eea{\end{eqnarray}}
\def\etal{{et al.}\thinspace}

\begin{document}

\title
[Suppression of thermal conduction in non-cooling flow clusters
]
{Suppression of thermal conduction in non-cooling flow clusters}

\author[Biman B. Nath]
{Biman B. Nath\\
Raman Research Institute, Bangalore 560080, India\\
(biman@rri.res.in)
}
\maketitle

\begin{abstract}
{
Recent X-ray observations have revealed a universal temperature
profile of the intracluster gas of non-cooling flow clusters which is 
flat for $r \le 0.2 r_{180}$. Numerical simulations, however, obtain
a steeper temperature profile in the inner region.
We study the effect of thermal conduction on the intracluster gas
in non-cooling flow clusters
in light of these observations, using the steep temperature profiles
obtained by authors of numerical simulations. We find that given $10^{10}$ yr
for the intracluster gas to evolve,
thermal conduction should be suppressed from
the classical value by a factor $\sim 10^{-3}$ in order to explain the
observations.
}
\end{abstract}


\section{Introduction}
The role of thermal conduction in the intracluster medium has long remained
uncertain. For clusters of galaxies which show signs
of cooling flow, it has been invoked in the past (Bertschinger \& Meiksin
1986; David, Hughes \& Tucker 1992) to allow multiphase cooling flows to exist,
although the degree of conduction was assumed to be 
suppressed below the classical
value (Binney \& Cowie 1981), possibly as a result of tangled magnetic
fields (e.g. Fabian 1994). In view of the recent observations of cooling 
flows, several authors have revived the idea of thermal conduction,
albeit with a suppression factor of only a few (Narayan \& Medvedev 2001;
Voigt \etal 2002). 

Concurrently, there have been two important developments in regard to the
temperature profile of non-cooling flow clusters. On one hand, 
observations with
{\it BeppoSAX} have revealed a universal temperature profile, normalised
by the emission weighted temperature of the whole cluster, which shows
flattening within $r \le 0.2 \, r_{180}$ 
(Molendi \& de Grandi 2001). On the other
hand, numerical simulations obtain a universal temperature profile that
is somewhat steeper in this inner region (Loken \etal 2002). 
One obvious physical mechanism that can flatten the temperature profile and
that has not been included in the simulation is thermal conduction
(Suginohara \& Ostriker 1998; Molendi \& de Grandi 2002; Loken \etal 2002).

Recently Loeb (2002) have argued on the basis of comparison of conduction
timescale with the age of clusters that thermal conduction should be suppressed
at least by a factor $\sim 0.15$ in order not to allow drastic cooling of the
gas in the cluster cores (of non-cooling flow clusters). 
Ettori \& Fabian (2000) argued that the observed temperature jumps in the 
cluster A2142 require the thermal conduction across cold fronts should be
suppressed by a factor larger than $\sim 100$. Vikhlinin \etal (2001)
also argued in favour of a suppression factor of $\sim 100$. 
Even in the bulk of the
gas, Markevitch \etal (2002) pointed out that a suppression factor larger than
$\sim 10$ is needed to explain the existence of small scale inhomogeneities
in the temperature profile of some clusters.

The physical mechanism for and the degree of suppression are
 still unclear though.
The proposed mechanisms include tangled magnetic field (Tribble 1989),
plasma instabilities (Pistinner, Levinson \& Eichler 1996, and references
therein). In the case of suppression by fluctuations in the
magnetic field, Chandran \& Cowley
(1998) found a suppression factor of order $10^{-2}\hbox{--}10^{-3}$. 
Pistinner, Levinson \& Eichler (1996) also advocated an inhibition factor of
order $\sim 10^{-3}$ as a result of electromagnetic instabilities. Recently
Narayan \& Medvedev (2001) extended the analysis of Chandran \& Cowley (1998) 
adding more wavevectors to the
fluctuation spectrum of magnetic fields and found a suppression factor of 
order $\sim 5$.

In this paper, we calculate the time evolution of the
 temperature profile of the intracluster
gas in non-cooling flow clusters in the presence of thermal conduction, with
various degrees of suppression, assuming quasi-hydrostatic equilibrium,
 and constrain the suppression factor by
comparing the resulting temperature profile with the {\it BeppoSAX} data.

We first discuss the temperature and density profiles of the intracluster
medium and the role of thermal conduction in \S 2. We then compare the
conduction timescale with the age of clusters in \S 3. We present the main
result of our work in \S 4.
Throughout the paper, we use the $\Lambda$CDM cosmological model, with
$\Omega_m=0.3$, $\Omega_{\Lambda}=0.7$ and $h=0.65$.

\section{Heat conduction and the intracluster gas}
Heat is conducted along the gradient of the electron temperature in a plasma.
Thermal conduction is termed unsaturated if the mean free path  of electrons
is smaller than the scale length of the temperature gradient. It is 
said to `saturate' otherwise. In the case of unsaturated thermal conduction,
 the heat flux is given by,
\be
{\mathbf Q}=-\kappa \nabla T_e \,,
\ee
where the thermal conductivity of a hydrogen plasma is given by (Spitzer 1962),
\be
\kappa_{sp}=1.8 \times 10^{-5} T_e^{5/2} (\ln \Lambda)_e ^{-1} {\rm erg} \, 
{\rm s}^{-1} \, {\rm cm}^{-1} \, {\rm K}^{-1} \,,
\label{eq:sp}
\ee
where the Coulomb logarithm is (Sarazin 1986) (with $n_e$ as the electron
density),
\be
(\ln \Lambda)_e=37.8+\ln \Bigr [ \Bigl ({T \over 10^8 \, {\rm K}} \Bigr ) \Bigl ( 
{n_e \over 10^{-3} \, {\rm cm}^{-3}} \Bigl )^{-1/2} \Bigr ] \,.
\ee

In the case of intracluster gas, we will find that the scale length of the
temperature gradient for the relevant temperature profile is larger than the
mean free path of electrons ($\sim  23 \, {\rm kpc} (T/10^8 \, K)^2 (n_e/10^{-3} \,
{\rm cm}^{-3})^{-1}$ (Sarazin 1986)) in general. 
This means that the above mentioned expression 
for unsaturated heat flux is relevant here. In fact, with the suppression
of thermal conduction, as is discussed below, the mean free path is 
smaller than the above value, 
bolstering the case for unsaturated heating due to thermal conduction.

Thermal conduction will tend to transport heat from the hotter (usually the inner)
regions of the intracluster gas to the colder (usually outer) parts. If the speed
of the ensuing gas flow is assumed to be slow, or, equivalently, if the gas density profile is
 assumed to be in an approximate stationary state, then one can write,
\be
{3 \over 2} {\rho_g (r) k_B \over \mu m_p} {\partial T(r,t) \over \partial t}
=-\nabla \cdot {\mathbf Q}=\nabla \cdot (\kappa \nabla T(r,t)) \,,
\label{eq:cond}
\ee
where $\rho_g (r), k_B, \mu,m_p$ are the gas density, Boltzmann constant, mean
molecular weight and the proton mass respectively.

\subsection{Temperature profile}
Recently, Molendi \& de Grandi (2002) have used the observations of
ten non-cooling flow clusters with {\it BeppoSAX} to derive a universal temperature
profile of the intracluster gas. They have normalised the profiles by the emission
weighted temperature of the cluster, and plotted the profiles as a function of
$r/r_{180}$, where $r_{180}$ is the radius where the mean overdensity of the
cluster is $180$ times that of the ambient density. The profiles are characterised 
by an isothermal core that extends to $r\sim 0.2r_{180}$, beyond which the temperature
declines in the outer region.

Molendi \& de Grandi (2002) have compared their observed temperature profile with
those predicted by several numerical simulations. Most of the predicted profiles 
are flatter than the observed profile in the outer region (Evrard \etal 1996;
Eke \etal 1998; Bialek \etal 2001). The high resolution simulation by
Frenk \etal (1999) did predict a temperature profile that matches well the observed
profile in the outer regions. Recently, Loken \etal (2002) have performed 
a numerical simulation with the highest resolution to date, and determined a universal
temperature profile that matches the observed profile of Molendi \& de Grandi (2002)
very well in the outer region ($r > 0.2 r_{180}$). They found a fit to their
universal profile that is given by (with $\langle T \rangle$ as the emission
weighted temperature), which we use as the initial temperature profile,
\be
T_i(r)={1.33 \langle T_i \rangle \over (1.+{1.5 r \over r_{vir}})^{1.6}}\,.
\ee

\subsection{Gas and dark matter density profile}
We assume the gas to be in a quasi-hydrostatic equilibrium
state as the temperature profile of the gas evolves. The initial density 
profile was calculated using the background dark matter density profile of
Navarro, Frenk \& White (1996) assuming hydrostatic equilibrium. The
dark matter density profile is then given by,
\be
\rho_{dm}(r)=\rho_s {1 \over (r/r_s) (1+r/r_s)^2} \,,
\ee
where $\rho_s$ is a normalizing density parameter. 
The characteristic radius $r_s$ is related
to the virial radius $r_{vir}$ by the `concentration parameter' ($c$), as
\be
c \equiv {r_{vir} \over r_{s}} \,.
\ee
The total mass of the cluster is assumed to be the mass inside its
virial radius.
The virial radius is calculated in the spherical collapse model to be,
\be
r_{vir}=\Bigl [ {M_{vir} \over (4 \pi /3 )
\Delta_c (z) \rho_c(z) } \Bigr ]^{1/3} \,,
\ee
where $\Delta_c(z)$ is the 
 spherical overdensity of
the virialized halo within $r_{vir}$ at $z$ in the units of the critical
density of the universe $\rho_c(z)$. For our adopted cosmological model
$\Omega_m=0.3$ and $\Omega_\Lambda=0.7$, the value of $\Delta_c(z)$ is
$100$ (Komatsu \& Seljak 2002). The total dark
matter mass within a radius $r$ is
\be
M(\le r)=4 \pi \rho_s r_s^3 m(r/r_s) \,,
\label{eq:totm}
\ee
where,
$m(x)=\ln (1+x)-{x \over 1+x}
$.

We first determine the initial gas density profile corresponding to the
initial temperature profile as discussed above, assuming hydrostatic
equilibrium,
\be
{d \over dr}\Bigl ({\rho_i(r) k_B T_i(r) \over \mu m_p}\Bigr )=
-\rho_i(r) {G M(<r) \over r^2} 
\,,
\ee
where $\rho_i$ is the initial  gas mass density and $M(<r)$ is the total mass
within a radius $r$ as given by equation (\ref{eq:totm}). We normalize the
gas density by requiring the total gas mass (within $r_{vir}$) to be a fraction
$f_g=\Omega_b/\Omega_m=0.15$ of the total mass of the cluster, where we have
used $\Omega_b=0.02 h^{-2}$ as constrained by primordial
nucleosynthesis (Burles \& Tytler 1998).

\section{Comparison of time scales}
One can define a conduction time scale for a spherically symmetric cluster
 as (Sarazin 1986;
Suginohara \& Ostriker 1998; Loeb 2002),
\be
t_{cond}={3 \over 2} {n_e k_B T \over \vert \nabla \cdot (\kappa \nabla T) \vert}
={3 \over 2} {n_e k_B T \over \vert {1 \over r^2} {d \over dr} (r^2 \kappa 
{dT \over dr})\vert} \,.
\ee
This can be compared with the age of the cluster, which we assume to be $10^{10}$
yr (as in, e.g., Loeb 2002). Here the conduction coefficient $\kappa$ is supposed
to be a fraction $\eta$ of its classical value $\kappa_{sp}$ (equation(\ref{eq:sp})).
 In Figure 1 we show (by solid lines) this time scale for the case of $\eta=0.1$
for clusters of masses $5 \times 10^{14},10^{15}, 2 \times 10^{15}$ M$_{\odot}$,
as a function of $r/r_{vir}$, and where the dotted line denotes the age of the
clusters. We have used the initial temperature and density profiles as discussed
in the preceding section. The discontinuity in the solid lines occur as
a result of the modulus in the previous equation.
It is seen from the comparison of time scales with $10^{10}$ yr that
for $\eta=0.1$, thermal conduction plays an important role in determining the
temperature profile for $r/r_{vir} \le 0.1\hbox{--}0.2$ for the above range of
cluster masses. It could be argued that for the cluster temperature profiles to
flatten below $r/r_{vir}\le 0.2$ in $\sim 10^{10}$ yr,
the suppression factor for thermal conduction
should be $\le 0.1$.  

Loeb (2002) argued on the basis of a similar comparison---constraining $t_{cond}/
10^{10} {\rm yr} \le 0.3$ at $r_{180}$---that $\eta
\le 0.15 (T/10 \, {\rm keV})^{-3/2}$ in order for the cluster cores not to have
cooled substantially over the last Hubble time. We note here that the emission
weighted temperature of clusters is related to the total mass as $M \propto T^{3/2}$
(Evrard \etal 1996), so that the limit on $\eta$ scales as $M^{-1}$.

Although the comparison of timescales is a good indicator, a much better limit
on the suppression factor is expected from the calculation of the time evolution 
the temperature profile as a result of thermal conduction. For example, 
from Figure 1
it is seen that for $\eta =0.1$, say for a cluster with mass $5 \times 10^{14}$
M$_{\odot}$, $t_{cond}=10^{10}$ yr at $r/r_{vir} \sim 0.1$. This only means that the
temperature would drop at least by a factor $(1-e^{-1})$ at that radius. This would,
however, mean that the temperature profile would be flat within a radius not
simply equal to $r/r_{vir}\sim0.1$ but considerably larger than this. In other words,
in reality the suppression factor could be much smaller than that obtained by
a simple comparison of time scales.

\begin{figure}
\centerline{
\epsfxsize=0.4\textwidth
\epsfbox{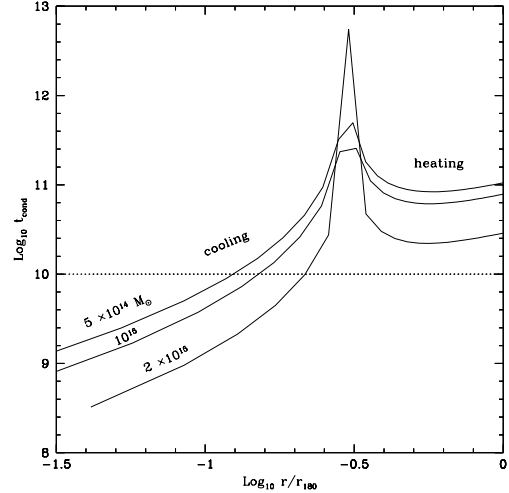}
}
{\vskip-3mm}
\caption{
Conduction time scale is compared to the age of the cluster of $10^{10}$ yr,
for three clusters of masses $5 \times 10^{14},10^{15},2 \times 10^{15}$ 
M$_{\odot}$ for a suppression factor of $\eta=0.1$. 
The regions of heating and cooling of the gas are marked. The dotted line
indicates the timescale of $10^{10}$ yr that relates to the cluster age. 
}
\end{figure}

\section{Evolution of temperature profile}
We next determine the evolution of the temperature profile using equation
(\ref{eq:cond}) with varying degrees of suppression of the value of $\kappa$.
For numerical integration of the equation with Crank-Nicholson method, 
it is helpful to change the variable
from $T$ to $T^{7/2}$ because $\kappa \propto T^{5/2}$ (equation(\ref{eq:sp})).
The most natural boundary condition is to keep the temperature at $r_{vir}$
a constant, where the effect of thermal conduction would be minimum. We
have therefore integrated equation(\ref{eq:cond}), with the boundary 
condition that the temperature at $r_{vir}$ is  a constant.

To facilitate comparison with the {\it BeppoSAX} temperature profile, which
is normalised by the emission weighted temperature, 
we have calculated the emission weighted temperature for the time-evolved
profiles, within $r_{2500}$ as,
\be
\langle T \rangle =
{\int_0^{r_{2500}} 4 \pi r^2 n(r)^2 \epsilon_{0.5-10} T(r) dr \over
\int_0^{r_{2500}} 4 \pi r^2 n(r)^2 \epsilon_{0.5-10} dr} \,,
\label{eq:weight}
\ee
where $n$ represents gas particle density and $\epsilon_{0.5-10}$
denotes the emissivity for a metallicity of $Z/Z_{\odot}=0.3$
relevant for the $0.5\hbox{--}10$ keV band, which we
calculated using the Raymond-Smith code. 
 We note that the initial profile
of Loken \etal (2002), the emission weighted temperature was calculated 
within a radius of $1 \, h^{-1}$ Mpc. For this profile, we 
have adopted the empirical scaling relation
of mass and emission weighted temperature of Finoguenov \etal (2001). 

\begin{figure}
\centerline{
\epsfxsize=0.4\textwidth
\epsfbox{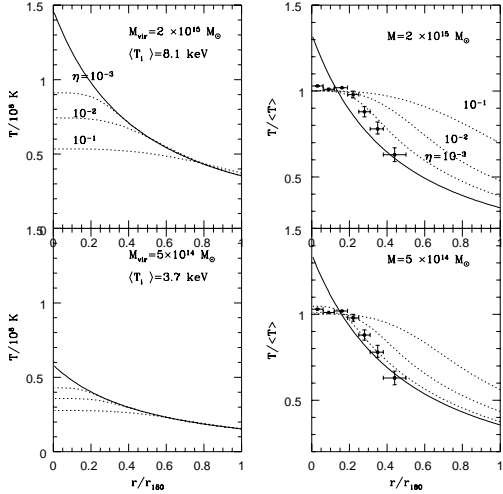}
}
{\vskip-3mm}
\caption{
Evolution of the temperature profile is shown for different suppression
factors of thermal conduction $\eta=10^{-1},10^{-2},10^{-3}$ (dotted lines),
beginning with the initial profile shown by the solid line, given
$10^{10}$ yr of evolution, for $M=2\times 10^{15}$ M${\odot}$ (upper
panels) and $M=5\times 10^{14}$ M$_{\odot}$ (lower panels). The left panels plot 
temperature against $r/r_{180}$ and the right panels plot 
the temperature normalized
by the corresponding emission weighted temperature.
}
\end{figure}

We plot in Figure 2 the results of numerically integrating equation
(\ref{eq:cond}) beginning with the initial temperature profile of
Loken \etal (2002) and the corresponding density profile, computed
on the basis of hydrostatic equilibrium. 
 The left panels show the temperature profile at the
end of $10^{10}$ yr for $\eta=10^{-1},10^{-2},10^{-3}$ (dotted lines)
and the right panels show the same for temperature normalized by
the corresponding emission weighted temperature. We plot the profiles
for $M=2\times 10^{15}$ and $5\times 10^{14}$ M$_{\odot}$. The
mass range is chosen to probe clusters with $\langle T \rangle$
in the range of $\sim 4\hbox{--}10$ keV.

The figures show that given an age of clusters of order $\sim 10^{10}$ yr,
a suppression factor for the thermal conduction that is $\eta \gg 10^{-3}$
produces temperature profiles that are flat for a large fraction
of the inner region and that are inconsistent with {\it BeppoSAX} observations. 
The discrepancies for a given suppression factor become larger for 
larger clusters, as expected since thermal conduction depends on $T^{5/2}$.
It is, however, seen that a value of $\eta \sim 10^{-3}$ results in a 
temperature profile that is consistent with the {\it BeppoSAX} data, almost
independent of the cluster mass.

\section{Discussion}
\subsection{Luminosity evolution}
We can also attempt to constrain the suppression factor from the fact that 
a substantial evolution in the temperature profile of a 
cluster over cosmological time scale would also change its X-ray luminosity
by a large amount in that time scale. Recent observations, however, find
that the X-ray luminosity of clusters scale as $L_x \propto (1+z)^{1.5}$
which is expected from clusters collapsing at various cosmological epochs
(see, e.g., Arnaud \etal (2002)). As $L_x \propto \rho^2 R^3 T^{1/2}$,
then with $T \propto (\rho R^3)/R$ and $\rho \propto (1+z)^3$, one finds
a scaling of $L_x$ at a given temperature with redshift as $(1+z)^{1.5}$.
Vikhlinin \etal (2002) found with {\it CHANDRA}
that for a sample of 22 clusters with $z>0.4$, the X-ray luminosity
for a fixed temperature scaled approximately as $(1+z)^{1.5 \pm 0.3}$.
If thermal conduction changed the temperature substantially then these
observations could be at variance with these simple cosmological scalings.

For the purpose of illustrating the point, consider a cluster of a given
mass collapsing at $z=0.5$, with an initial profile similar to the one
we considered in the previous section (and a corresponding density
profile from the consideration of hydrostatic equilibrium, taking into account
the changes in the dark matter potential and in the fraction 
$\Omega_b/\Omega_m$, appropriate for a collapse
redshift of $z=0.5$).
If we now let this cluster passively evolve
with thermal conduction at work, then we could track its evolution in
the $L_x-T$ space.
Vikhlinin \etal (2002) have shown that the data correlate well if the
emission weighted temperature is plotted against $L_x \times (1+z)^{1.5}$
(their Figure 2b). In Figure 3, we have plotted the evolutionary track of two
representative clusters (of masses $10^{15}$ and $2\times 10^{15}$ M$_{\odot}$)
from $z=0.5$ to $z=0.0$, for the same parameters. We
have calculated $L_x$ within a radius
of $2$ Mpc and the emission weighted temperature within a shell of $0.5\hbox{--}
1$ Mpc, as Vikhlinin \etal (2002) have done. We also show the data points from
Vikhlinin \etal (2002) as well as their best fit by the dotted line.

The exact calibration of $L_x$ and emission weighted temperature is not
important here, as the initial temperature (density) profile has been chosen
for the purpose of illustration, not taking into account how the `universal'
temperature profile could evolve in redshift. The important points to
note in this figure are the extent and direction of the evolutionary tracks.
The tracks show how
thermal evolution could introduce scatter in this plot. As Voit (2000)
has argued the difference between the collpase redshift of a cluster and
the redshift at which it is observed can introduce scatter in the plots
relating the X-ray parameters. 

A small
value of thermal conduction coefficient would only move it 
along the $L_x (1+z)^{1.5}$
axis, mainly because of the redshift factor. 
Although thermal conduction only conducts heat from the inner to the
outer part of the cluster, and on average the gas temperature does not change,
but since X-ray emission depends on the density, temperatures determined from
X-ray emission would be lowered by thermal conduction.
A large value of the coefficient would result in a track in which both
$L_x$ and temperature are changed with time. The data at present 
still have a large scatter and it is not possible to constrain the value of
thermal conduction from this, but it might be possible to do so in the near
future, if the uncertainties in the temperature measurements decrease the
errorbars in the $T$ direction.

Similar evolutionary tracks could also be used for other X-ray parameters,
e.g., in the $M-T$ space. The scatter in the
data for total mass and the emission weighted
temperature, especially for high redshift clusters, such as in Schindler (1999),
however
is too large to allow any useful constraint on thermal conduction.
As Loeb (2002) has pointed out  a substantial evolution in the temperature
profile would also introduce a large scatter in the  measurements of
cluster abundance. While these constraints would become better with more
data in the future, we have obtained a much stronger limit from the 
measurement of the temperature profile as described in the previous section.

\begin{figure}
\centerline{
\epsfxsize=0.4\textwidth
\epsfbox{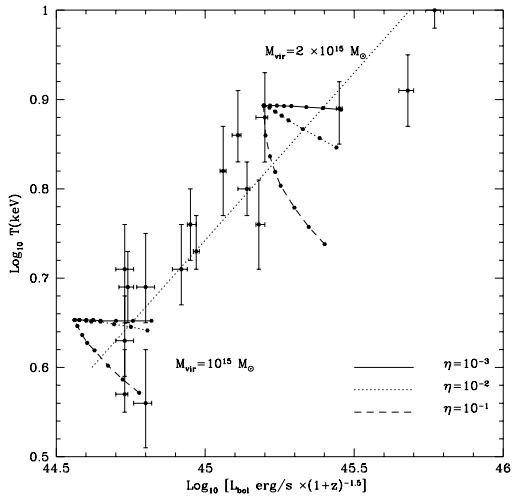}
}
{\vskip-3mm}
\caption{
Evolution of a cluster collapsing at $z=0.5$ in the space of 
$L_x \times (1+z)^{-1.5}$ and emission weighted temperature is shown for
two clusters with $M=10^{15}$ and $2\times 10^{15}$ M$_{\odot}$. Solid, dotted
and dashed lines show the evolution for $\eta=10^{-3}, 10^{-2}, 10^{-1}$.
Filled circles along the tracks indicate the luminosity and temperature
at $z=0.5,0.45,0.4,0.35,0.2,0.1,0.0$; the point of convergence of tracks
refers to $z=0.5$.
Data points and the dotted line is from Vikhlinin \etal (2002).
}
\end{figure}

\subsection{Limitations}
Finally, we wish to reiterate the basic assumptions made in our calculations
to remind ourselves of the limitations of the results. The gas was
assumed to be in an approximate steady state condition. We have also
done our calculations relaxing this assumption somewhat, by requiring
the gas density profile to be in hydrostatic equilibrium at every time step.
This did not alter the results much, and we believe that our results
should be robust in this regard, as long as there is no large scale gas
flow. It should also be pointed out that we have assumed a single value
of the suppression factor for the whole of the intracluster gas, independent
of density or temperature.

We  assumed the intracluster gas to evolve passively in time, without any
cooling and heating sources acting in it. It is possible that heating
sources, such as active galactic nuclei (B\"ohringer \etal 2002), 
which could be more prevalent in the
inner region, could compensate the cooling due to thermal conduction in this
region to some extent. When radiative cooling is dominant in the central 
region, as in cooling flow clusters, the direction of heating due to thermal
conduction may reverse from being outward to inward. The central region would
then heat instead of getting cooled as discussed earlier in the paper.
As a matter of fact, it has been speculated that radiative cooling
may change the structure of the magnetic field and decrease the suppression
of thermal conduction in the central region (Soker \& Sarazin 1990).
It is also possible that both AGN heating and radiative cooling  are
important in cooling flow clusters;
Ruszkowski \& Begelman (2002) found that heating due to
thermal conduction suppressed only to the extent of $\eta \sim 0.1$,
combined with AGN heating, may explain the
recent observations of cooling flow clusters.

We have also implicitly assumed the suppression factor to be homogenous throughout
the intracluster gas, which may not always be valid, especially when cooling is
important (see above). Fabian, Voigt and Morris (2002) have discussed the
possibility that conduction is stronger in the central region than in the
outer radii owing to the change in the structure of magnetic fields brought
about by cooling in the centre.

\section{Summary}
We have determined the time evolution of the temperature profile
as a result of thermal conduction in the limit that the density profile
is in steady state, beginning with temperature profiles obtained from
numerical simulations (with thermal conduction). We have constrained
the suppression factor for the thermal conduction to be of order
$\le 10^{-3}$ by comparing the final
temparature profile (after $10^{10}$ yr) with the observed profiles.
We have also calculated the corresponding  evolution of the X-ray
luminosity of clusters at a given temperature, but found the 
observed data too scattered to allow any useful constraint on 
thermal conduction.

\bigskip
\noindent
{\bf Acknowledgement}
It is a pleasure to thank Dipankar Bhattacharya, David Eichler and Abraham
Loeb for stimulating discussions and comments on the manuscript. The 
comments of the anonymous referee have also improved the paper, and they
are acknowledged with thanks.

\end{document}